\documentclass[%
    reprint,
    aps,
    superscriptaddress,
    showpacs,
    prb,
    floatfix,
    bibnotes,
    amssymb,
    amsfonts,
]{revtex4-2}

\usepackage[utf8]{inputenc}
\usepackage[T1]{fontenc}
\usepackage{amsmath,amsthm,mathtools,mathrsfs,physics}
\usepackage{graphicx}
\usepackage{subfig}
\usepackage{hyperref}
\usepackage{color}
\usepackage{siunitx}
\usepackage{booktabs}
\usepackage{multirow}
\graphicspath{{img/}}
\usepackage{xcolor}
\definecolor{darkgreen}{rgb}{0.0, 0.5, 0.0}
\DeclareSIUnit\angstrom{\text{Å}}
\DeclareSIUnit\rydberg{Ry}
\DeclareUnicodeCharacter{2212}{-}

\usepackage{braket}
\usepackage{bbold}

\begin{document}
\title{Disorder-induced conducting edges on Kagomé lattice}

\author{A. Chmeruk}
\affiliation{Theoretische Physik III, Center for Electronic Correlations and Magnetism, Institute of Physics, University of Augsburg, 86135 Augsburg, Germany}
\affiliation{Augsburg Center for Innovative Technologies (ACIT), University of Augsburg, 86135 Augsburg, Germany}
\author{D. Jones}
\affiliation{Theoretische Physik III, Center for Electronic Correlations and Magnetism, Institute of Physics, University of Augsburg, 86135 Augsburg, Germany}
\affiliation{Augsburg Center for Innovative Technologies (ACIT), University of Augsburg, 86135 Augsburg, Germany}
\author{L. Chioncel}
\affiliation{Theoretische Physik III, Center for Electronic Correlations and Magnetism, Institute of Physics, University of Augsburg, 86135 Augsburg, Germany}
\affiliation{Augsburg Center for Innovative Technologies (ACIT), University of Augsburg, 86135 Augsburg, Germany}

\date{\today}

\begin{abstract}
Within a cluster extension of the coherent potential approximation, disorder averaging generates a non-local self-energy that renormalizes both diagonal and off-diagonal hopping terms of the non-interacting Kagome-lattice Hamiltonian.
%
These renormalizations 
(of both nearest- and next-nearest-neighbor hopping amplitudes) drive the system between two topologically trivial insulating states through an intermediate gapless phase characterized by conducting edge modes over a broad range of impurity concentrations.
Our results demonstrate that multiple-scattering effects alone can generate emergent effective spin-orbit interactions and qualitatively modify the edge spectrum of disordered Kagome systems.
%
%
\end{abstract}

\maketitle
 
\section{Introduction}

The electronic structure of disordered solids is a mature yet rapidly evolving field, driven by continuing advances in effective-medium theories~\cite{sove.67,ve.ki.68,shib.71,yo.mo.73,el.kr.74, le.ra.85}, cluster methods~\cite{tsuk.69,duca.74,ja.kr.01,rowl.06}, and first-principles techniques capable of treating disorder~\cite{gyor.72,te.gy.78,eb.ko.11,ko.lo.11,vi.sk.00,vito.01}, including electronic correlations~\cite{ch.vi.03,mi.ch.05, os.vi.17,os.vi.18}, and localization sometimes on equal footing~\cite{te.zh.17,os.zh.20,we.zh.21,jani.26}.

Effective medium theories~\cite{sove.67,ve.ki.68,shib.71,yo.mo.73, el.kr.74,le.ra.85} are powerful approaches in modeling the effects of disorder in substitutional alloys
by considering 
%
a disorder-averaged effective medium that restores translational invariance. 
At the level of a non-interacting disordered model, randomness can enter the Hamiltonian matrix in various ways. The simplest case corresponds to having on-site energies randomly distributed on a well-defined lattice with the hopping elements being independent of a particular configuration. 
This is the case of diagonal disorder. 
A more complicated type of disorder includes the randomness (beyond the on-site energies) in the hopping integrals (amplitudes) as well. This can be addressed by the Blackman-Esterling-Berk (BEB)~\cite{bl.es.71,este.75} reformulation 
%
of the coherent potential approximation (CPA) and its non-local extensions~\cite{goni.92}.
%
%
%

An effective medium theory, particularly the CPA, can be formulated within an iterative framework. This approach, implemented through a sequence of Dyson-like equations, enables one to monitor the development of the effective medium by tracking the renormalization of its constituent building blocks. In the simplest case of a single atom per unit cell, the on-site energies acquire a complex, energy-dependent self-energy that both shifts the energy levels and provides the electronic states with a finite lifetime. In contrast, other parameters governing the band structure, such as the interatomic hopping integrals, remain unchanged. Although the CPA self-energy remains local in the sense that it is defined as a single-site (lattice-point) quantity, it nevertheless incorporates a degree of non-local information through the 
environment of
atomic sites associated with that lattice point.
The CPA framework can be generalized in a straightforward manner to incorporate nearest-neighbor hoppings. This is achieved by considering a finite cluster of atomic sites as the unit cell, which by construction includes the intersite hoppings. 
Referring to the vast literature on the topic of cluster-methods, most significant extensions belong to the family of molecular CPA~\cite{tsuk.69,duca.74}, a description in the real space, and the dynamical cluster approximation~\cite{ja.kr.01,rowl.06} formulated in momentum space. 
Formulating the effective theory in terms of these clusters makes it possible to systematically account for the effects of disorder 
beyond the single-site description. 

In this work we will focus on computing disordered averages of local observables for a two-dimensional Kagome lattice Hamiltonian. 
The family of materials containing two-dimensional Kagome lattices exhibits a rich spectrum of low-energy electronic states that display diverse manifestations of band non-trivial topology: Chern insulators, Weyl semimetals~(WSM), topological superconductors, and quantum spin liquids~\cite{th.ko.82, ha.ka.10,yi.li.22, bo.na.19, ar.me.18}.
It is largely accepted that weak disorder is generally compatible with - and often preserves - the topological features of electronic structures~\cite{shoucheng_2011, schnyder_2016}, provided that the disorder is not strong enough to trigger non-trivial metal insulator transitions 
or induce Anderson localization.
Crucial quantities, such as averages of the one-particle Green's function (averaged density of states) does not become critical across the Anderson localization as no single-particle gap is formed. In principle localization would require to study the full statistical distribution of such averages. 

To study the effects of disorder, we consider the Kagome lattice in a strip geometry, where a finite-size cluster is periodically repeated along one direction. Disorder is introduced by replacing host clusters with impurity clusters, resulting in an effective chain of disordered clusters. This geometry allows us to track the evolution of both bulk and edge properties and, in particular, to monitor the modification of the edge states, which are directly connected to the topological properties of the system.
%
%
%


The manuscript is organized as follows. Sec.~\ref{sec:gen_comp} presents the theoretical setup and the computational details. The effective low-energy model of the real-space cluster used in our computation is presented in Sec.~\ref{sec:H_eff}. It is based on the known Hamiltonian for Kagome lattice~\cite{gu.fr.09}. Here, it is formulated in a way which is most suitable for the application of cluster CPA.
We provide in Sec.~\ref{sec:cluster_CPA} a pedagogical review of our real-space embedding method based on the molecular CPA-ideology.  We give a detailed summary of the self-consistent embedding approach in the locator formalism. 
In Sec.~\ref{sec:results} we present the numerical results. We observe a formation of a gapless state for a wide range of concentrations. The CPA self-energy is analyzed as well. 
Sec.~\ref{sec:discus} discusses the  possible mechanisms and relation to the known disorder-induced topological insulators. 
Finally, we discuss also the possibility of the extension of the present embedding technique into  available \emph{ab initio} methods to make it possible to address realistic systems. 

\section{General theory and computational details}
\label{sec:gen_comp}
In the following section Sec.~\ref{sec:H_eff} we review the low-energy model of the two-dimensional Kagome layer, formulated in a tight-binding language~\cite{gu.fr.09} which we adapt to a suitable form for our embedding cluster approach. This model is particularly fitting for a description of chemical doping, for instance with a concentration of dopants ($c_B$) introduced into the host material ($c_A$). Fluctuations in the on-site energies $\epsilon_i = \epsilon_{A/B}$ are referred to as diagonal disorder.   
A natural framework for discussing various real-space approaches to the problem of disorder is the locator expansion~\cite{ande.58,el.kr.74,       ma.ja.05}. 
We make an extensive use of it in presenting our method in Sec.~\ref{sec:cluster_CPA}.

\subsection{Cluster Hamiltonian for the two-dimensional Kagomé lattice and the edge spectrum}
\label{sec:H_eff}
Consider first the non-interacting limit of the Kagomé lattice -  Hamiltonian~\cite{gu.fr.09}:
\begin{align}
    \nonumber H = &\sum_{i,\sigma}\ket{i,\sigma}\varepsilon_{i} \bra{i,\sigma} +  
    \sum_{\langle i, j,\sigma \rangle}\ket{i,\sigma}t_{i,j} \bra{j,\sigma} +\\
    &\sum_{\langle \langle i, j, \sigma_{\alpha}, \sigma_{\beta} \rangle \rangle }\ket{i,\sigma_{\alpha}}\lambda_{i,j}(\mathbf{d}_i \times \mathbf{d}_j)_{{\alpha},{\beta}} \bra{j,\sigma_{\beta}}.
\label{eq:kag_ham}
\end{align}
%
%
Here, $i,j$ are the site indices. Single brackets $\langle...\rangle$ denote nearest-neighbor (NN) coupling and the double brackets $\langle\langle...\rangle\rangle$ denote the next-nearest neighbor (NNN) coupling. 
The last term describes the Kane-Mele like spin-orbit coupling (SOC)~\cite{ka.me.05}, with $\alpha,\beta$ label spinor components, and the cross product $\mathbf{d}_i \times \mathbf{d}_j$ is determined by the two bond vectors connecting the NNN sites through their common nearest neighbor. Given the strictly two-dimensional Kagome lattice: $\mathbf{d}_i \times \mathbf{d}_j = \pm |\mathbf{d}_i \times \mathbf{d}_j|\hat{z}$ the SOC operator reduces to: $\pm |\mathbf{d}_i \times \mathbf{d}_j|\sigma_z$, where the $\pm$ is a geometric-sign, positive if rotating $\mathbf{d}_i$ into $\mathbf{d}_j$ is counterclockwise and negative otherwise. 
%
The spin-up and spin-down sectors remain decoupled and acquire opposite phases in the NNN hopping amplitudes
similar to the Haldane model~\cite{gu.fr.09, haldane_1988}. Thus, the last term in the Bloch Kagome-lattice Hamiltonian is transformed accordingly: 
\begin{align*}
\sum_{\langle \langle i, j, \alpha,\beta \rangle \rangle }&\ket{i,\sigma_{\alpha}} \lambda_{i,j} (\mathbf{d}_i \times \mathbf{d}_j)_{{\alpha},{\beta}} \bra{j,\sigma_{\beta}} \rightarrow   \\
&\sum_{\langle \langle i, j,\sigma \rangle \rangle}\ket{i, \sigma}e^{i\phi_{i,j}} \lambda_{i,j} \bra{j, \sigma} 
\end{align*}

In order to access the one-dimensional edge modes, one needs to perform a dimensional reduction~\cite{shoucheng_2011}.
By reduction, the two-dimensional Bloch Hamiltonian Eq.~\eqref{eq:kag_ham} is transformed into a strip geometry that is finite along $\mathbf{a}_1$ and translationally invariant along $\mathbf{a}_2$. 
The translational symmetry along $\mathbf{a}_2$ allows the crystal momentum $\mathbf{k}_2$ to remain a good quantum number, while the finite extent along $\mathbf{a}_1$ gives rise to physical boundaries supporting topological edge states.
%
 
We begin by choosing a tiling of the lattice with identical stripes (clusters) each containing $N_c$-sites.
A site inside the cluster is denoted by 
at $\boldsymbol{\tau}_l$, with $l=\overline{1,N_c}$, while a lattice vector $\mathbf{r}_l$ in the original lattice can now be decomposed as: $\mathbf{r}_l =\mathbf{R} + \mathbf{\tau}_l$, where $\mathbf{R}$ labels the cluster. 
The original lattice Fourier transform can be rewritten using a reciprocal-space decomposition for the inter-cluster contribution and 
an internal cluster-Fourier transform. 
Under this conventions ~\cite{Vanderbilt_2018} the general real-space tight-biding Hamiltonian connecting a site $l$ belonging to a cluster located at $\mathbf{R}$ with a site $m$ of a cluster at $\mathbf{R}^\prime$: 
\begin{align*}
H=\sum_{R,R^\prime}\sum_{lm} c^\dagger_{R,l} t_{l,m}(R-R^\prime) c^{}_{R^\prime,m } 
\end{align*}
is re-written in the mixed real-space (intra-cluster) and momentum (inter-cluster) representation as:
\begin{align}
\label{hamk}
H &=  \sum_{\tilde{k}} \sum_{l,m} c^\dagger_{l}(\tilde{\mathbf{k}}) \left[ \sum_{\mathbf{T}} H_{l,m}( \mathbf{T}) e^{-i \tilde{\mathbf{k}} (\mathbf{T} + \boldsymbol{\tau}_m - \boldsymbol{\tau}_l)  }\right]  c^{}_{m }(\tilde{\mathbf{k}}),
\end{align}
with $\tilde{\mathbf{k}}$ belonging to the reduced Brillouin zone of the superlattice.
The (relative) vectors $\mathbf{T}=\mathbf{R}-\mathbf{R}^\prime$ correspond to translations along the periodic direction.
Finally, the inter-cluster Fourier transformed Hamiltonian is:
\begin{align}
\label{eq:H_lm}
H_{lm}(\tilde{\textbf{k}}) &= \sum_{\textbf{T}}
H_{lm}(\mathbf{T})
\mathrm{e}^{-i\tilde{\textbf{k}}(\textbf{T} + \boldsymbol{\tau}_m - \boldsymbol{\tau}_l)}
\end{align}
One can restrict the $\mathbf{T}$-sums to, e.g., $\mathbf{T} =0$ in the above equation and write:
\begin{align}\label{eq:H_C}
H_{lm}^C(\tilde{\textbf{k}}) &= H_{lm}^C
\mathrm{e}^{-i\tilde{\textbf{k}}(\boldsymbol{\tau}_m - \boldsymbol{\tau}_l)} ,  
\end{align}
with $H_{lm}^C$ being the real-space Hamiltonian within the cluster.
Further on, for $\mathbf{T} = \pm \mathbf{a}_2$, we introduce the notation:
\begin{align}\label{eq:H_CC'}
    W_{lm}(\tilde{\textbf{k}}) &= 
    \sum_{\textbf{T}=\{\pm\textbf{a}_2 \}}\mathrm{e}^{i\tilde{\textbf{k}}(\textbf{T} + \boldsymbol{\tau}_m - \boldsymbol{\tau}_l)}
    H_{lm}^{CC^\prime}(\textbf{T}) \ , 
\end{align}
where $H_{lm}^{CC^\prime}(\textbf{T})$ represents the inter-cluster couplings.
Putting Eq.~\eqref{eq:H_C} and~\eqref{eq:H_CC'} the full lattice Hamiltonian is then written in the approximation to truncate $\mathbf{T}$ to the nearest neighbors: $H \approx H^{C}_{lm}(\tilde{\textbf{k}}) + W_{lm}(\tilde{\textbf{k}})$. 
For completeness of notations we repeat the real-space form of the cluster Hamiltonian (a matrix of size $N_c \times N_c$): 
\begin{align*}
    \nonumber H^{C}_{lm} = \sum_{l=1,N_c}\ket{l}\varepsilon_{l} \bra{l} \ \delta_{lm}
&+\sum_{\langle l, m,\sigma \rangle_{C}}\ket{l, \sigma}t_{l,m} \bra{m, \sigma} \\
 &+ \sum_{\langle \langle l, m,\sigma \rangle \rangle_C}\ket{l, \sigma}\tilde{\lambda}_{l,m} \bra{m, \sigma} 
\end{align*}
and introduce capital letters denoting the inter-cluster hopping $T_{lm}$ and the effective inter-cluster SOC $\Lambda_{lm}$.   
The coupling between the clusters has than the corresponding form:
\begin{align*}
    \nonumber H^{CC^{\prime}}_{lm} =& \sum_{ \langle l, m,\sigma  \rangle _{CC^{\prime}}}\ket{l, \sigma}T_{l,m} \bra{m,\sigma} \\
    & +
    \sum_{\langle \langle l, m,\sigma \rangle \rangle _{CC^{\prime}}}\ket{l, \sigma}\tilde{\Lambda}_{l,m} \bra{m, \sigma}
\end{align*}
Its Fourier transform produces $W_{lm}(\tilde{\textbf{k}})$ explicitly present into the CPA selfconsistency loops.
Note that the complex hopping phase $\phi_{lm}$ taken to be $\pm\frac{\pi}{2}$ for spin-up and spin-down channels, is included in the redefined $\tilde{\lambda}$ and $\tilde{\Lambda}$.

\begin{figure*}
\includegraphics[width=1.0\textwidth]{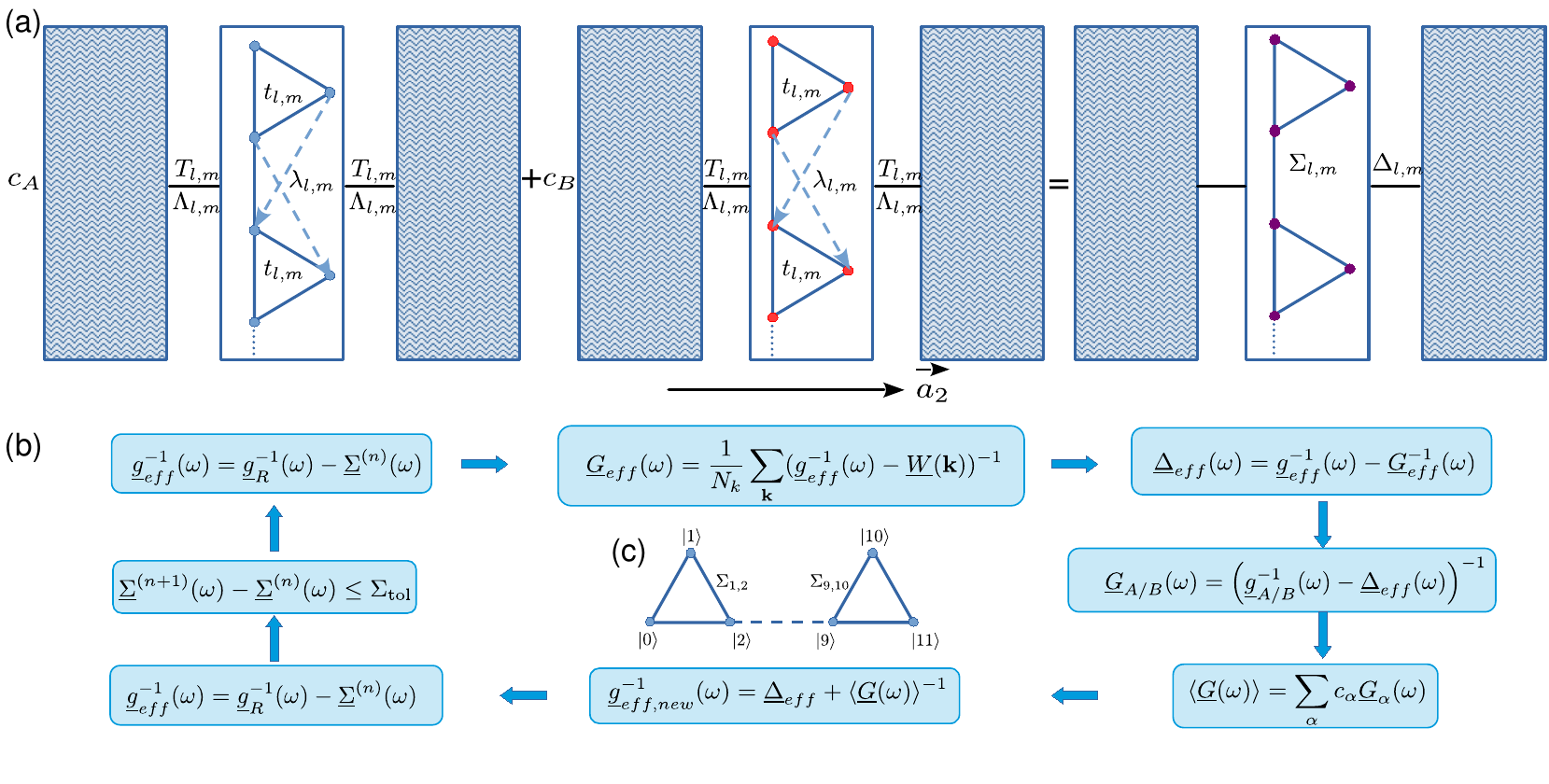}
\caption{(a) Schematic representation of the Kagome-lattice tiling with $N_c=6$ (the stripe here contains 6 sites) periodized along $\textbf{a}_{2}$ direction. 
The solid lines represent intra-cluster NN hopping parameters $t_{i,j}$ and the NNN $\lambda_{i,j}$ complex SOC couplings are depicted with dashed lines. The black line between the clusters depict inter-cluster couplings ($T_{ij}, \Lambda_{ij}$). (b) Flow diagram of the computational self-consistent embedding scheme Eq.~\eqref{eq:g_eff} $\rightarrow$ (c) Enumeration of the basis sites inside the cluster and the corresponding off-diagonal components of the self-energy ~\eqref{eq:Sigma_new} 
\label{fig:latt}}
\end{figure*}

This representation is particularly convenient for the subsequent cluster CPA formulation, since it naturally separates intracluster and intercluster hopping processes.
All parameters $\epsilon_l$ the local on-site energy, the inter-site hoppings $t_{lm},~T_{lm}$ and the strength of the spin-orbit coupling $\lambda_{lm},~\Lambda_{lm}$ can be treated as random variables. However, we will restrict ourselves to the case in which $\epsilon_l$ is a random variable with a probability distribution of a binary mixture.  

\subsection{Self-consistent coherent potential approximation within the locator-interactor formalism}
\label{sec:cluster_CPA}
With the notation of the previous section, we start by defining the bare locator $\underline{g}_R(\omega)$ of an isolated cluster at position $R$:
\begin{align*}
    \underline{g}_R(\omega) = \left[ \omega \underline{\mathbb{1}} - 
     \begin{pmatrix}
    \epsilon_1 &  & \\
    & \ddots & \\
    &  & \epsilon_{N_c} 
    \end{pmatrix}  - \tilde{\underline{t}}  \right]^{-1}
\end{align*}
where the diagonals contain the random-site energies $\epsilon_l$ and all intra-cluster hoppings are collected into $\tilde{t}$:
\begin{align*}
    \tilde{\underline{t}} = \sum_{\langle i, j,\sigma \rangle_{C}}\ket{i, \sigma}t_{i,j} \bra{j, \sigma} + \sum_{\langle \langle i, j,\sigma \rangle \rangle_C}\ket{i, \sigma}\tilde{\lambda}_{i,j} \bra{j, \sigma} 
\end{align*}
The stripe of type $A/B$ corresponds to clusters with on-site energies $\epsilon_l=\epsilon_{A/B}$ for all sites $l=1,N_c$. The underline here is the notation for the matrix of size $N_c \times N_c$.
One introduce at this point an effective locator:
\begin{equation}
\label{eq:g_eff}
\underline{g}_{eff}^{-1}(\omega) = \underline{g}_R^{-1}(\omega) - \underline{\Sigma}(\omega)   \ , 
\end{equation}
from which the effective Green's function is computed:
\begin{align}
    \underline{G}_{eff}(\omega) = \frac{1}{N_k} \sum_{\mathbf{k}} (\underline{g}_{eff}^{-1}(\omega) - \underline{W}(\mathbf{k}))^{-1}.
\end{align}
The corresponding effective interactor is:
\begin{align}
     \underline{\Delta}_{eff}(\omega) =  \underline{g}_{eff}^{-1}(\omega) -  \underline{G}_{eff}^{-1}(\omega).
\end{align}
The impurity GF is obtained by placing the impurity cluster locators into the effective medium:
\begin{align}
     \underline{G}_{A/B}(\omega) &= \left( \underline{g}_{A/B}^{-1}(\omega) - \underline{\Delta}_{eff}(\omega)\right)^{-1} 
\end{align}
with $\underline{g}_{A/B}^{-1}(\omega)$ being the $\underline{g}^{-1}(\omega)$ for which all on-site energies  $\epsilon_l$ are of the same type $\epsilon_{A/B} $. The CPA averaged Green's function has the form:
\begin{align}
    \langle \underline{G}(\omega)  \rangle =  \sum_{\alpha} c_\alpha \underline{G}_\alpha (\omega),
\end{align}
which permits the definition of a new locator:
\begin{align}
 \underline{g}_{eff,\text{new}}^{-1}(\omega) = \underline{\Delta}_{eff} + \langle \underline{G}(\omega)  \rangle^{-1}.   
\end{align}
Finally, a new self-energy can be computed as:
\begin{align}\label{eq:Sigma_new}
  \underline{\Sigma}_{\text{new}}(\omega) = \underline{g}_{R}^{-1}(\omega) - \underline{g}_{eff,\text{new}}^{-1}(\omega) 
\end{align}
The coherent self-energy:
\begin{equation}
    \underline{\Sigma}(\omega) =
    \begin{pmatrix}
    \Sigma_{1,1} & \Sigma_{1,2} & \dots & \Sigma_{1,N_c} \\
    \Sigma_{2,1} & \Sigma_{2,2} & \dots & \Sigma_{2,N_c} \\
    \vdots & \vdots & \ddots & \vdots \\
    \Sigma_{N_c,1} & \Sigma_{N_c,2} & \dots & \Sigma_{N_c,N_c}
    \end{pmatrix}
\label{eq:se_matrix}
\end{equation}
contains diagonal disorder corrections and off-diagonal corrections describing correlations within the cluster. The new coherent locator is then used to recalculate the effective medium GF. The self-consistency is achieved when the average GF and the effective medium GF are the same. Equivalently, the condition can be formulated in terms of the self-energy as:
\begin{align*}
    || \Sigma^{(n+1)} - \Sigma^{(n)} || \le 10^{-6},
\end{align*}
where $n$ is the number of iterations. The effect of substitutional disorder is fully captured in the renormalized locator $\underline{g}_{eff}$. Its self-energy, i.e., the interactor that connects it to the bare (host) lattice locator can be extracted using Eq.~\ref{eq:g_eff}. 
Various quantities, such as the $\textbf{k}$-resolved spectral function, are readily available:
\begin{equation}
    A(\textbf{k},\omega) = -\frac{1}{\pi}\mathrm{Im}[\mathrm{Tr} \  \underline{G}(\mathbf{k}, \omega)],
\label{akw}
\end{equation}
where
\begin{equation}
    \underline{G}(\mathbf{k}, \omega) = [\underline{g}_{0}^{-1}(\omega) - \underline{\Sigma}(\omega) - \underline{W}(\mathbf{k})]^{-1}.
\end{equation}
%
The total density of states (DOS) is found by $\textbf{k}-$integrating and tracing the average GF:
\begin{equation}
    N(\omega) = -\frac{1}{\pi}\int\mathrm{Im}[\mathrm{Tr} \  \underline{G}(\mathbf{k}, \omega)]d\textbf{k}.
\label{dos}
\end{equation}
It will also prove useful to define a partial density of states (PDOS) of the bulk and the edge. This is done by limiting the trace of the GF to components that belong, respectively, to the bulk and the edge, namely:
\begin{align}
    N_{bulk}(\omega)= -\frac{1}{\pi}\int\mathrm{Im}\left[ \sum_{l=2,...,9}  G_{ll}(\mathbf{k}, \omega)\right] d\textbf{k}, \label{eq:N_bulk}\\
    N_{edge}(\omega)= -\frac{1}{\pi}\int\mathrm{Im}\left[ \sum_{l=0,1,10,11}  G_{ll}(\mathbf{k}, \omega)\right] d\textbf{k}, \label{eq:N_edge}\\
\end{align}
where the partial trace is written explicitly. The enumeration of the basis states is shown in Fig.~\ref{fig:latt}(c). The number of electrons is given by:
\begin{equation}
    \int_{-\infty}^{E_{F}}\int_{BZ}A({\textbf{k},\omega}) \ d\mathbf{k} \ d\omega = n_{E}.
\label{el_count}
\end{equation}

\subsection{Computational details}

The cluster method presented above that computes averages, formulated in real space treats exactly fluctuations stemming from the impurity scattering up to the size of the cluster ($N_c$). Our numerical computation indicated a favorable convergence of properties with $N_c$ as it is a controlled approach of the solution with increasing cluster size.

We present results for a strip-like geometry composed of $N_c=12$ sites within a cluster   along the $\textbf{a}_1$ direction while along $\textbf{a}_2$ periodic boundary conditions are used.
A given disorder configuration is characterized by the concentration of host stripes ($c_A$) and the concentration of impurity stripes ($c_B$), which obeys the condition $c_A+c_B =1$. Note that the host stripes have the on-site energies on every site of the cluster $\epsilon_A$, while the impurity stripes have on-site energies $\epsilon_B$.
We define the disorder strength $\delta = \varepsilon_A - \varepsilon_B$ and consider the weak SOC limit for which the intra-cluster $\lambda_{lm} = 0.15$ is of the same magnitude to the inter-cluster SOC: $\Lambda_{lm} = \lambda_{lm}$. We fix $\varepsilon_{A}=1$ and vary $\varepsilon_{B}$. Since the disorder strength $\delta$ is the main parameter determining the formation of the effective medium, a particular choice of $\varepsilon_A$ and $\varepsilon_B$ is arbitrary. Moreover, the CPA is symmetric with respect to the choice of host/impurity and this selection does not affect the results. It is convenient to set $\varepsilon_{A}=1$, which sets the Fermi level at the zero energy in the absence of disorder at $\frac{1}{3}$ filling. Then, the clusters with $\varepsilon_{B}$ are taken as impurities.


The strip geometry retains translational symmetry only along the $\mathbf{b}_2$ direction. 
As a result, the Hamiltonian, Eq.~\eqref{eq:H_lm}, depends on a single Bloch momentum ($\mathbf{k}_2 \equiv \tilde{\mathbf{k}}$), whereas the finite direction is represented explicitly in real space. 
Accordingly, the Brillouin-zone integration is performed only along the reciprocal lattice vector ($\mathbf{b}_2$). 
A uniform mesh of ($\mathbf{k}_2$)-points is employed over the reducible Brillouin zone, assigning identical weights to all sampling points.


We use the self-energy defined in Eq.~\ref{eq:g_eff} as the convergence criterion, which is equivalent to requiring that the average and effective-medium GFs coincide. Convergence is achieved when the self-energy changes by less than 10$^{-6}$. Additionally, convergence is considered achieved only if the resulting GF contains no negative values in the spectrum. In the present case this is not an issue, however it can be for large disorder strength. The calculations are performed on the complex energy grid $\omega = E + i\eta$. The smearing $\eta$ is taken to be $10^{-4}$. In certain cases of large disorder strength, the CPA convergence can become very slow or not achievable at all. In such cases, mixing of the $(n)$-th and $(n+1)$-th iterations of the self-energy can be used to speed up convergence. The Fermi level is found via Eq.~\ref{el_count}. This integral is highly sensitive to the choice of \textbf{k}-mesh, energy mesh and the smearing. For the accurate determination we use 128 \textbf{k}-points, 2701 energy points and the smearing of 10$^{-4}$, respectively.

We would like to stress that the choice of the real hopping $t$ and SOC $\lambda$ define the bare lattice and do not ever change. The cluster self-energy, however, is generally non-diagonal and can acquire finite renormalizations for NN and NNN transfer integrals in the self-consistency loop. 

\section{Results}
\label{sec:results}

The following sections present the electronic properties of the disordered Kagome strip calculated within the cluster-CPA approach. 
We begin with the momentum-resolved spectral function, which reveals the disorder-induced modifications of the band structure, and subsequently examine the corresponding DOS. 
Finally, we analyze the local and intersite components of the self-energy to characterize the spatial structure of the effective disorder potential and its possible connection to the appearance of the gapless edge modes.



\subsection{Bloch spectral function $A(\textbf{k},E)$}

\begin{figure}[h!]
\includegraphics[width=0.5\textwidth]{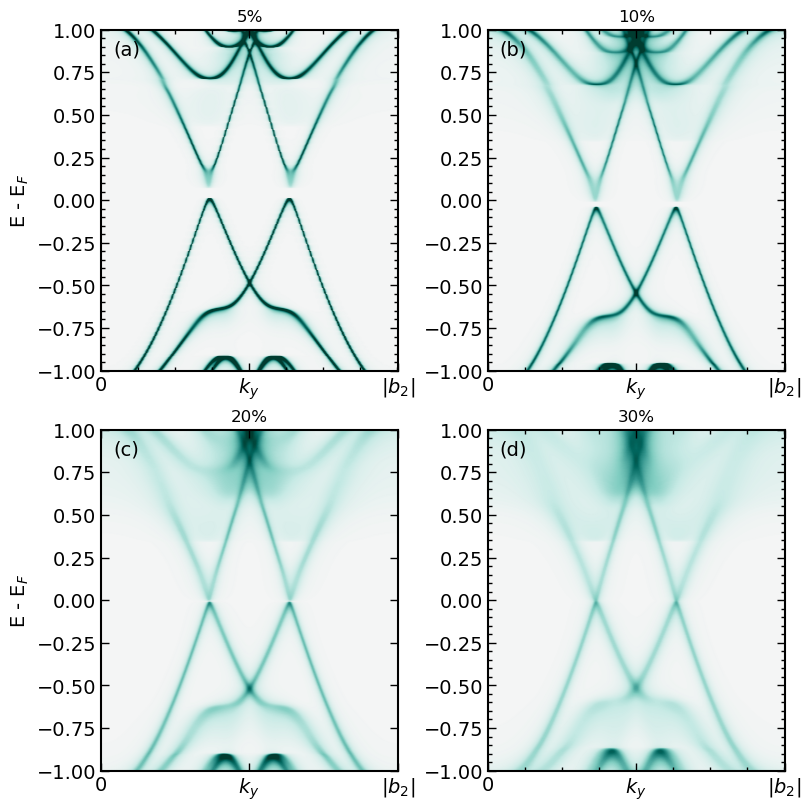}
\caption{The Bloch spectral function with $\delta=0.6$ with varying concentrations up to 30\% of $B$-type clusters. The gap closes at around 20 \% . The smearing is $\eta=10^{-4}$}
\label{Bloch_high}
\end{figure}
In Fig.~\ref{Bloch_high}, the Bloch spectral function is shown for a fixed  disorder strength $\delta=0.6$ and varying concentration. The impurity clusters/stripes have the on-site energies $\varepsilon_B$ for all cluster sites $N_c=12$. Since the SOC chosen is not strong enough these states remain gapped in both limits with $\underline{\epsilon}_{A/B}$.
A clear gap is visible for a concentrations of 5\% and 10\% of $B$-type clusters into the $A$-type host, Fig.~\ref{Bloch_high}(a-b). As the concentration of impurity type-$B$ clusters is increased, the gap starts to fill with spectral weight. 
If the disorder strength is large enough, there will be a critical concentration at which the gap will close. In the present case this occurs at around 20 \% of impurity concentration, Fig.~\ref{Bloch_high}(c). 
Increasing the concentration further preserves the previously formed metallic state, Fig.~\ref{Bloch_high}(d). Since the CPA is symmetric w.r.t the choice of the host, the gapless state will remain until the host concentration is reduced to 20 \%. At this point, the gap reopens again.

\subsection{DOS: gap filling mechanism}
\label{sec:DOS}
For the momentum-resolved spectral function, we considered $\delta=0.6$, where the modification of the band structure and the evolution of the in-gap states are clearly resolved
(see Fig~\ref{Bloch_high}). For the DOS, a slightly smaller disorder strength ($\delta=0.5$), is considered. 
This is used to provide a clearer visualization of the disorder-induced redistribution of spectral weight and the associated gap evolution. 
Note that these values are selected for illustrative purposes and do not represent different disorder regimes.

%
%
%
%
%

\begin{figure}[t!]
    \centering
    \includegraphics[width=0.5\textwidth]{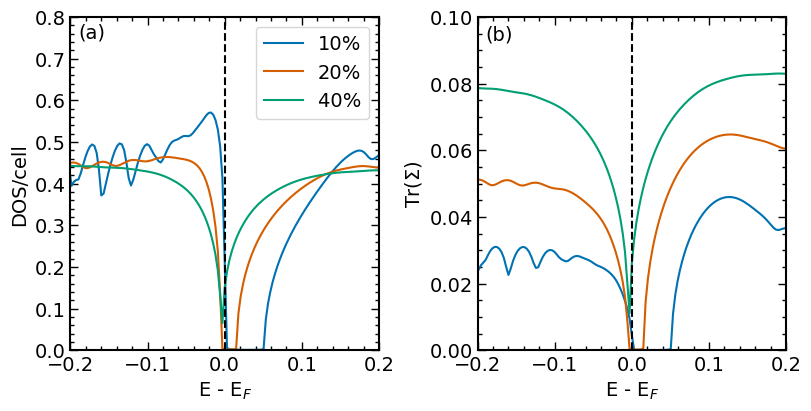}
    \caption{The evolution of the total DOS with concentration at $\delta =0.5$ reveals the gap closure at the concentration of 40\%.}
    \label{TDOS}
\end{figure}

In Fig.~\ref{TDOS}, we correlate the evolution of the total density of states with the concentration-dependent self-energy at fixed disorder strength, $\delta =0.5$. As the concentration is increased from 10\% to 40\%, the insulating gap is progressively reduced and is essentially closed at 40\% for the present parameters.

The corresponding evolution of the cluster self-energy is shown in Fig.~\ref{TDOS}(b). 
The increase in the magnitude of $Tr \ \Sigma$ with concentration reflects the increasing strength of the disorder-induced renormalization of the effective medium. 
The simultaneous reduction of the spectral gap in Fig.~\ref{TDOS}(a) therefore results from the combined effect of the real and imaginary parts of the self-energy: the real part shifts the disorder-renormalized band-edges relative to the Fermi energy, while the imaginary part introduces a finite spectral broadening and associated tails. As the concentration is increased, these effects progressively transfer spectral weight toward the gap region until the band edge reaches the Fermi energy and the system enters a gapless spectral regime.

\begin{figure}[t!]
    \centering
    \includegraphics[width=0.5\textwidth]{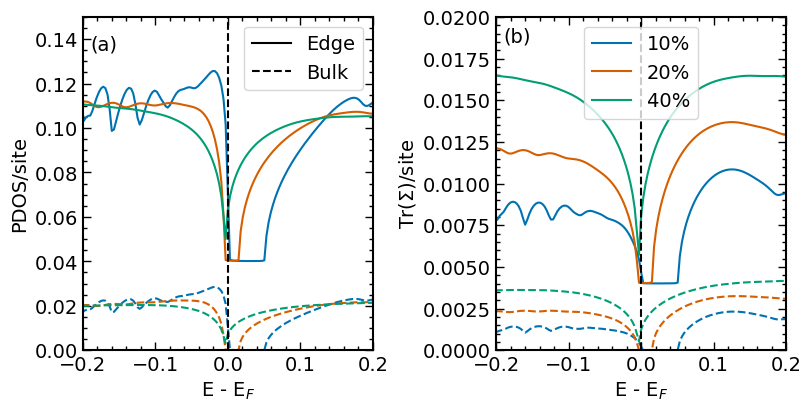}
    \caption{The evolution of the PDOS with concentration at $\delta = 0.5$. To highlight the gap a different offsets for edge-PDOS (0.04) and $\Sigma$ (0.004) are introduced}
    \label{LDOS}
\end{figure}

The cluster CPA approach allows to follow also the evolution of the PDOS for both edge and bulk contributions,~Fig.~\ref{LDOS}.  
The edge and bulk partial densities of states are defined according to Eqs.~\eqref{eq:N_bulk} and~\eqref{eq:N_edge}, respectively, and are normalized per site, i.e., by the number of components entering the corresponding trace. The edge contribution is vertically shifted by (0.04) for clarity. This normalization allows a direct comparison of the spectral weight associated with the two spatial regions.

The edge contribution is consistently larger than the bulk contribution, demonstrating that the low-energy spectral weight is predominantly concentrated near the boundaries. 
The evolution with concentration shows that the suppression of $N(E)$ around the Fermi energy is modified for both the edge and bulk states. 
The progressive reduction of the gap with increasing concentration, already seen in Fig.~\ref{TDOS}, is accompanied by the appearance of spectral weight near ($E_F$) for both edge and bulk states. 
Therefore the gap closure is not exclusively an edge effect, although the magnitude of the resulting low-energy spectral weight is strongly enhanced at the boundaries. 
The unequal edge contributions is a consequence of the inequivalent terminations of the two boundaries of the finite Kagome strip.

The corresponding concentration dependence of the cluster self-energy is shown in Fig.~\ref{LDOS}(b). 
The edge and bulk components of ($\mathrm{Tr} \ \Sigma$) exhibit the same qualitative trend as the PDOS, with the edge contribution being substantially larger on a per-site basis. Increasing the concentration from 10\% to 40\% enhances the magnitude of the disorder-induced self-energy and modifies its energy dependence near the Fermi level. 
The simultaneous evolution of the self-energy and the PDOS indicates that the concentration-driven gap closure in the total DOS is associated with the increasing disorder-induced renormalization of the effective medium. 
In particular, the stronger self-energy response at the boundaries is consistent with the enhanced edge spectral weight observed in the PDOS.

Note that the larger edge contribution to ($\mathrm{Tr} \  \Sigma$) should not be interpreted as a larger bare hopping or spin-orbit coupling at the boundary. 
Rather, it reflects the spatial dependence of the effective disorder self-energy generated by the cluster-CPA in the finite strip geometry. 
The bare Hamiltonian parameters remain fixed; the spatial variation arises from the inequivalent local environments experienced by sites near the boundary and in the bulk.


%
\subsection{Off-diagonal 
cluster self-energies $\Sigma_{lm}(\omega)$}

The locator-interactor formulation of the cluster CPA allows to track the development of the self-consistently determined effective medium cluster. Its properties are contained within the cluster self-energy  $\Sigma_{lm}(\omega)$. The off-diagonal components of the cluster self-energy have previously been interpreted as self-consistently found transfer integrals between different sites inside the cluster~\cite{go.ga.78, goni.92}. 

The off-diagonal components of $\Sigma_{lm}(\omega)$ reveal how the effective disorder medium evolves on the nearest- and next-nearest-neighbor bonds adjacent to the boundary. This allows us to formulate a microscopic (though semi-quantitative) origin of the DOS and PDOS results discussed in Sec.~\ref{sec:DOS}. 
Figs.~\ref{TDOS} and~\ref{LDOS} established that increasing impurity concentration drives the system from a gapped to a gapless spectral regime and that the low-energy spectral weight is strongly enhanced at the edges.


\begin{figure}[t!]
    \includegraphics[width=0.5\textwidth]{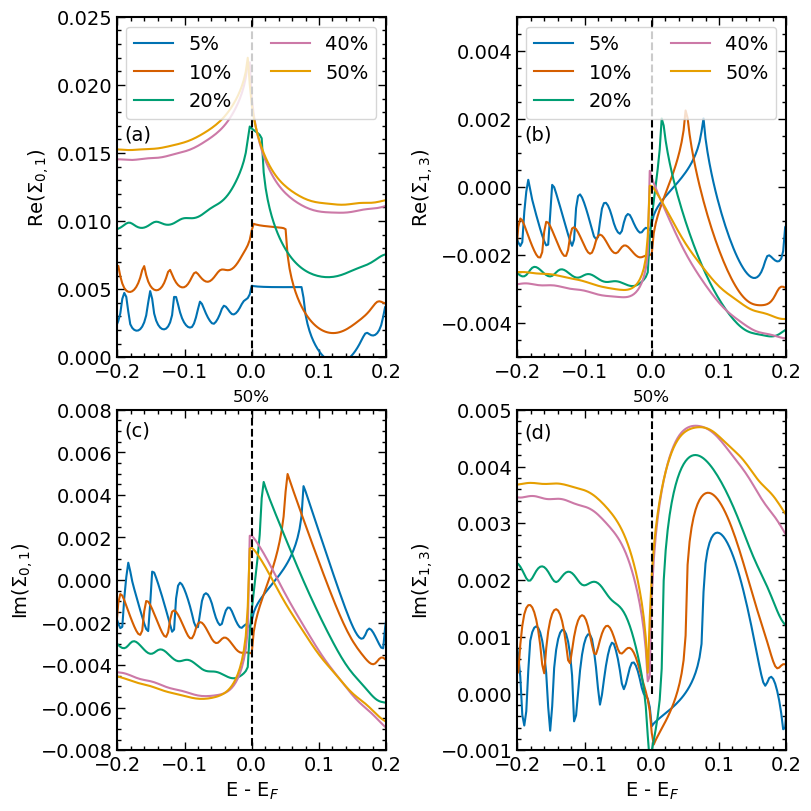}
    \caption{Off-diagonal components of the cluster self-energies for $\delta=0.5$ near the edge.
    }
\label{se_off}
\end{figure}

Fig.~\ref{se_off} shows representative off-diagonal components of the cluster self-energy for bonds located in the boundary region of the Kagome strip. The real and imaginary parts of the nearest-neighbor component $\Sigma_{01}$, connecting the edge site $0$ to its nearest neighbor are shown in Fig.~\ref{se_off}(a, c), respectively. Panels (b) and (d) display the corresponding real and imaginary parts of the NNN component $\Sigma_{13}$, for a bond within the edge region. Together with the total DOS and the edge- and bulk-resolved PDOS, these quantities provide a microscopic view of the concentration-dependent renormalization of the disordered effective medium in the region close to the edge.

The real part of the NN component, Re$(\Sigma_{01})$, exhibits a systematic increase in magnitude with increasing disorder concentration. At low concentrations, the self-energy is relatively small and displays pronounced oscillatory structure below the Fermi energy. With increasing concentration, these oscillations are progressively suppressed and the energy dependence becomes smoother, while a pronounced variation develops in the vicinity of $E_F$. Since the off-diagonal self-energy enters the effective propagator together with the bare hopping matrix elements, its real part represents an energy-dependent nonlocal renormalization of the electronic propagation between the two sites. The concentration dependence of Re$(\Sigma_{01})$ is therefore consistent with the progressive modification of the band edges and the reduction of the spectral gap observed in the total DOS of Fig.~\ref{TDOS}.

The NNN component $\Sigma_{13}$ exhibits a similar continuous evolution with concentration, although its magnitude and energy dependence differ from those of the NN component. In particular, the real part of $\Sigma_{13}$ remains small compared to the corresponding NN contribution, while the imaginary part is of comparable magnitude to its NN counterpart. Both contributions develop a pronounced energy dependence around and above the Fermi level. The presence of a finite NNN component demonstrates that the cluster treatment generates a genuinely nonlocal disorder self-energy extending beyond NN bonds. Importantly, the bare NN and NNN hoppings (or SOC) remain fixed parameters of the underlying Kagome Hamiltonian; the concentration- and energy-dependent contribution shown here is instead a self-energy correction generated by the disordered effective medium. 

The imaginary parts of the off-diagonal self-energies characterize the absorptive component of this nonlocal scattering correction. Their evolution with concentration is smooth and does not exhibit the appearance of a distinct additional structure that could be associated with a new scattering channel. Rather, the increasing magnitude and changing energy dependence of these components indicate a continuous modification of the disorder-induced scattering processes as the concentration is increased. In combination with the real parts, they modify both the position and the spectral broadening of the disorder-renormalized states.

These results complement the PDOS shown in Fig.~\ref{LDOS}. The substantially larger edge spectral weight is accompanied by pronounced nonlocal self-energy components on bonds located near the boundary. Nevertheless, the corresponding self-energy is not restricted to the edge: the cluster effective medium contains both edge and bulk contributions, consistent with the simultaneous evolution of the edge and bulk PDOS. The gap closure observed in the total DOS therefore results from the cumulative effect of the concentration-dependent disorder self-energy on the electronic spectrum, rather than from the appearance of a new hopping or SOC term.

Finally, the off-diagonal self-energies evolve continuously from the dilute-disorder regime, where $\Sigma_{lm} \xrightarrow[]{} 0$ as the concentration $c \rightarrow 0$, through the concentration range in which the DOS evolves from a gapped to a gapless spectrum. This behavior supports a picture in which the spectral evolution is driven by a continuous renormalization of the effective medium by disorder. Within the present calculation, the gap closure is therefore not accompanied by an abrupt onset of a distinct nonlocal scattering channel.

\begin{figure}[t!]
    \includegraphics[width=0.5\textwidth]{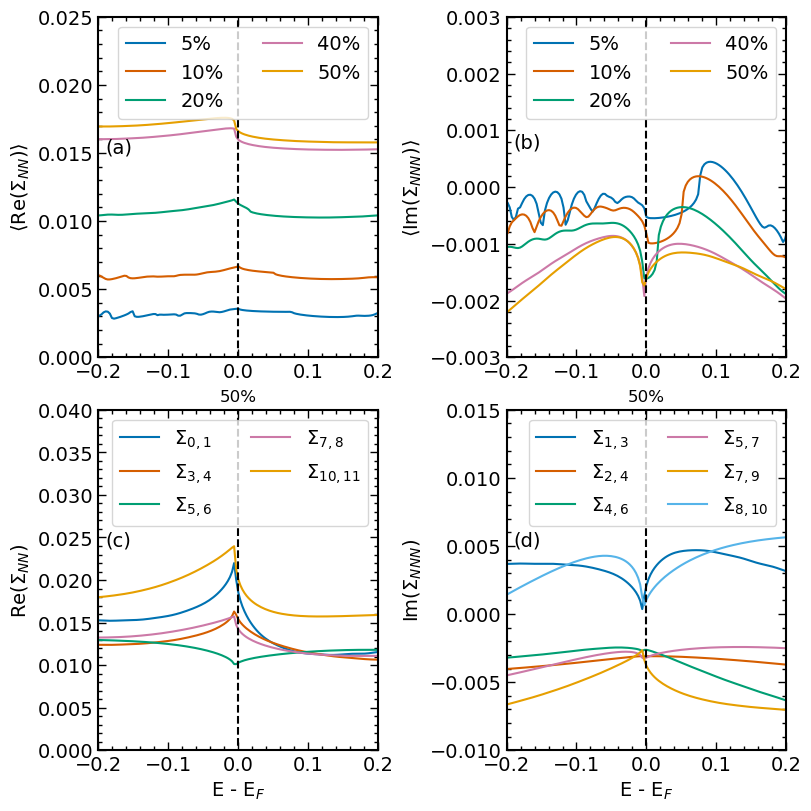}
    \caption{Off-diagonal components of the cluster self-energies for $\delta=0.5$. (a-b) Average magnitude of the real and imaginary parts of the NN and NNN components, respectively. (c-d) Selected individual components at 50\% concentration. Enumeration of the matrix elements is shown in Fig.~\ref{fig:latt}(c). 
    }
\label{se_off2}
\end{figure}

Fig.~\ref{se_off2} summarizes the concentration and bond dependence of the off-diagonal cluster self-energy. Panels (a) and (b) show the concentration evolution of the real and imaginary parts of the self-energy averaged over the NN and NNN components, respectively. The corresponding individual matrix elements at the highest concentration, c=50\%, are resolved in panels (c) and (d), providing information on the bond-to-bond variations within the cluster.

The averaged real part of the NN self-energy, Fig.~\ref{se_off2}(a), increases systematically in magnitude with concentration while retaining a relatively smooth energy dependence. 
In the dilute limit the off-diagonal self-energy is small, consistent with its interpretation as a disorder-induced nonlocal correction to the clean effective medium. Increasing the concentration enhances this correction and produces a pronounced energy dependence around the Fermi energy. The corresponding imaginary part exhibits a more structured energy dependence, including a pronounced variation around $E_F$ and a redistribution of spectral weight above the Fermi level,~Fig.~\ref{se_off2}(b). The evolution of the real and imaginary parts demonstrates that increasing disorder modifies both the dispersive and absorptive components of the nonlocal self-energy of the effective medium.

Panels (c) and (d) resolve the individual off-diagonal components for $c=50\%$, at which the effects of the self-energy are the most pronounced. The NN components have comparable magnitudes but exhibit distinct energy dependences, Fig.~\ref{se_off2}(c). In particular, some components develop pronounced structures close to $E_F$, whereas others remain relatively smooth. 
This demonstrates that the cluster self-energy is not characterized by a single universal NN matrix element. Instead, it retains information about the local geometry and the inequivalent environments of the bonds within the finite strip.

A similar bond dependence is observed for the NNN components, Fig.~\ref{se_off2}(d). The individual $\Sigma_{lm}$ elements differ in both magnitude and energy dependence, with some components changing sign or developing pronounced extrema over the energy range shown. Their non-equivalence further reflects the reduced translational and point-group symmetry associated with the strip geometry and its boundaries. Their concentration dependence and bond dependence provide a direct indication of the spatial structure of disorder scattering captured by the cluster CPA.


The averaged quantities in panels (a) and (b) therefore establish the systematic growth of nonlocal disorder effects with concentration, while panels (c) and (d) demonstrate that these effects are distributed non-uniformly among the individual bonds of the cluster. This provides a microscopic complement to the DOS and PDOS results: the concentration-driven evolution from a gapped to a gapless spectrum is accompanied by a continuous, spatially and bond-dependent renormalization of the effective medium.

\section{Discussion and Outlook}
\label{sec:discus}

\begin{figure*}
    \includegraphics[width=1.0\textwidth]{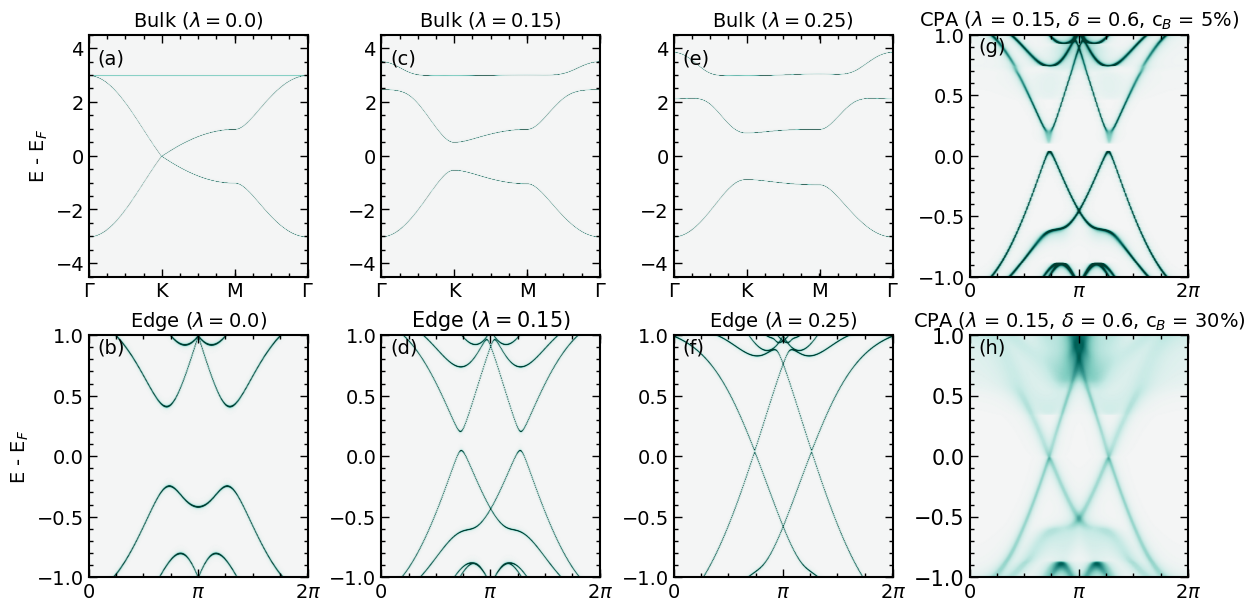}
    \caption{Comparison of the SOC driven band gap closure and the disorder induced one. (a-b) In case of no SOC, Kagome bulk is in semimetalic state with gapped edge states. (c-d) Turning on SOC opens the gap in the bulk and produces the edge states that start to close the edge gap. (e-f) Once the SOC is strong enough, the bulk is fully gapped and the edge spectrum exhibits spin-filtered linearly dispersing bands. (g-h) In case when SOC is not strong enough yet to close the gap in the host, introducing impurities may lead to the closing of the edge gap.}
\label{bulk_edge_cpa}
\end{figure*}

We studied the evolution of the electronic spectrum of the disordered Kagome strip as a function of impurity concentration at fixed disorder strength, $\delta=\epsilon_A-\epsilon_B$. 
For sufficiently strong disorder, increasing the impurity concentration progressively reduces the spectral gap and, within the range of parameters investigated, leads to a gapless regime. The cluster CPA provides access not only to the total density of states but also to its spatially resolved edge and bulk contributions and to the nonlocal components of the disorder self-energy.

The gap closure in the edge spectrum is highly reminiscent of the appearance of topologically protected states in Kagome systems upon turning on the SOC~\cite{gu.fr.09, bo.na.19}. The present calculation however, does not provide direct access to the bulk topological invariant of the corresponding two-dimensional system. The appearance of edge-enhanced gapless spectral weight should therefore not, by itself, be interpreted as evidence for a topological phase. In the absence of an explicit calculation of a bulk topological invariant, we cannot establish that the disorder-induced gapless states are the consequence of a topologically non-trivial bulk. Nevertheless, it is instructive to compare it with the SOC-driven emergence of the non-trivial edge states in the Kagome Hamiltonian. As shown in Fig.~\ref{bulk_edge_cpa}, the clean system exhibits a qualitatively different evolution when the SOC is varied. When SOC is turned off, the bulk spectrum remains gapless while the edge spectrum retains a finite gap, Fig.~\ref{bulk_edge_cpa}(a, b). Increasing the SOC strength opens a bulk gap and simultaneously modifies the edge spectrum, Fig.~\ref{bulk_edge_cpa}(c, d), while for sufficiently strong SOC the bulk becomes fully gapped and the edge spectrum develops the characteristic linearly dispersing states associated with the topological Kagome phase, Fig.~\ref{bulk_edge_cpa}(e, f).

The disordered system exhibits a qualitatively similar spectral evolution. For a host system whose SOC is insufficient to produce a fully gapless edge spectrum, increasing the concentration of impurities at sufficiently large $\delta$ can progressively reduce the edge gap and eventually produce gapless spectral weight, Fig.~\ref{bulk_edge_cpa}(g, h). The similarity between the clean SOC-driven evolution and the disorder-driven evolution suggests that both mechanisms can strongly modify the low-energy spectrum. It does not, however, establish that the disorder acts as an effective SOC interaction. Rather, the disorder enters through the energy- and concentration-dependent self-energy of the effective medium.
%
%
In this work, we have not been able to find a closed analytic expression for the critical disorder strength (and impurity concentration) at which the gapless modes emerge. It is clear, however, that it has to be \emph{at least} larger than the band gap of the host system.
%
%
%

The concentration dependence of the self-energy is smooth across the range in which the DOS evolves from a gapped to a gapless spectrum. This supports a picture in which the spectral evolution results from a continuous disorder-induced renormalization and broadening of the effective medium rather than from the abrupt onset of a distinct scattering channel. The real and imaginary parts of the self-energy contribute, respectively, to the dispersive renormalization and absorptive broadening of the electronic propagator, while their nonlocal components encode the spatial correlations of disorder scattering within the cluster.

The cluster CPA approach used in this work evidently leads to local symmetry breaking inside the cluster. The translational symmetry is only preserved at the level of superlattice. A highly asymmetric cluster self-energy makes it difficult to pinpoint exactly whether the emergent gapless modes are due to the NNN self-energies emulating emergent SOC-like terms, or other longer-range terms. It is possible to improve the current approach by restoring the full symmetry of the lattice. This improved approach is known as the dynamical cluster approximation~\cite{jarrell_2005}.

The effective medium approach was previously used to explain the emergence of the quantized conductance in disordered HgTe/CdTe quantum wells~\cite{Groth_2009, bern2006}. Using the self-consistent Born approximation~(SCBA), the authors showed that disorder-induced band inversion through the renormalization of the mass term is the driving mechanism behind the TAI transition. Recently, the self-consistent T-matrix approximation was employed to investigate how disorder drives a chirality transition in the Haldane model~\cite{Knolle_2025}. The resulting phase was dubbed as genuine topological Anderson insulator (GTAI). Importantly, the impurities considered in that study were intrinsically topologically non-trivial. Within our CPA framework, a similar state can emerge if the impurities are assumed to possess a different SOC from that of the host lattice. However, since the SOC takes the form of NNN hopping, this effectively introduces off-diagonal type of disorder, which is inherently absent in the CPA. The present framework can be extended to account for these effects using Blackman-Esterling-Berk approach, which can be formulated in a very similar iterative self-consistent scheme used in this work~\cite{bl.es.71}.

Including off-diagonal disorder is especially interesting for integrating it with the available \emph{ab initio} methods. The go-to method for modeling doped systems is to consider larger cells of specific disorder realizations. Although it allows to include local effects, this method is bound by the cell sizes and therefore only specific concentrations of dopants can be realized. The effective medium approach overcomes this limitation and offers a continuous range of impurity concentrations. The tight-binding formulation used in this work connects most naturally to the Wannerized Hamiltonians that can be extracted from \emph{ab initio} techniques~\cite{wan_cpa_2022, wan_cpa_2025}. The biggest challenge lies in determining the Wannier representation of the bare lattice.

In the present work, we have considered a disordered chain, where the appearance of gapless edge states is to be understood as a consequence of the disorder-induced modifications of the bulk spectrum rather than being incorporated explicitly into the effective medium description. A fully self-consistent treatment would require deriving a coupled set of self-consistency equations that capture the effects of disorder in the bulk and consistently communicate these modifications to the edge. Although this extension is, in principle, a natural generalization of the present framework, it is computationally significantly more demanding. We leave this analysis for future work

\begin{acknowledgments}
Financial support by the Deutsche Forschungsgemeinschaft (DFG, German Research Foundation) – TRR 360, project no. 492547816, subproject A5 is gratefully appreciated. 
\end{acknowledgments}

\appendix


\section{The CPA self-energy, locators and interactors in the propagator formalism}
It is well-known that the CPA and its cluster extensions can be formulated in either locator-interactor or propagator formalism. Both yield the same result. There is, however, a difference in how the self-energies are defined. The relation between the two can be shown in the following way. We will use capital letters for locator-related quantities and small ones for propagators: 
\begin{align}
    (\underline{g}_0^{-1} - \underline{\sigma} - \underline{\delta})^{-1} = \langle \underline{G}\rangle\\
    (\underline{G}_{0}^{-1} - \underline{\Sigma})^{-1} = \langle \underline{G}\rangle
\end{align}
The bare propagator $G_{0}$ can be evaluated as:
\begin{align}
    \underline{G}_{0} = \frac{1}{N_{k}}\sum_{\textbf{k}}(\underline{g}_{0}^{-1} - \underline{W}(\textbf{k}))^{-1}.
\end{align}
Through the Dyson equation, one can extract the interactor (or "self-energy") that connects it to the bare locator:
\begin{align}
    \underline{\Delta} = \underline{g}_{0}^{-1} - \underline{G}_{0}^{-1}.
\end{align}
Plugging it into (A2) we arrive at:
\begin{align}
    &(\underline{g}_0^{-1} - \underline{\sigma} - \underline{\delta})^{-1} = \langle \underline{G}\rangle\\
    &(\underline{g}_{0}^{-1} - \underline{\Sigma} - \underline{\Delta})^{-1} = \langle \underline{G}\rangle\\
    &\underline{\Sigma} = \underline{\sigma} + \underline{\delta} - \underline{\Delta}
\end{align}
In the present approach, both $\underline{\sigma}$ and $\underline{\delta}$ are modified in the self-consistency cycle, whereas in the propagator formalism only $\underline{\Sigma}$ is modified. Thus, by knowing $\underline{\sigma}$ and $\underline{\delta}$ one can always recover $\underline{\Sigma}$ but not the other way around.
\bibliographystyle{apsrev-title.bst}
\bibliography{sources.bib}

\end{document}